\documentclass[a4paper,12pt]{article}
\usepackage{graphicx}
\begin{document}
\newcommand{\beqn}{\begin{eqnarray}}
\newcommand{\eeqn}{\end{eqnarray}}
\newcommand{\ra}{\rightarrow}

\newcommand{\np}{Nucl.\,Phys.\,}
\newcommand{\pl}{Phys.\,Lett.\,}
\newcommand{\pr}{Phys.\,Rev.\,}
\newcommand{\prl}{Phys.\,Rev.\,Lett.\,}
\newcommand{\prep}{Phys.\,Rep.\,}
\newcommand{\nuclinst}{{\em Nucl.\ Instrum.\ Meth.\ }}
\newcommand{\annp}{{\em Ann.\ Phys.\ }}
\newcommand{\intjmp}{{\em Int.\ J.\ of Mod.\  Phys.\ }}


\newcommand{\mw}{M_{W}}
\newcommand{\mww}{M_{W}^{2}}
\newcommand{\mwmw}{M_{W}^{2}}

\newcommand{\mz}{M_{Z}}
\newcommand{\mzz}{M_{Z}^{2}}

\newcommand{\cw}{\cos\theta_W}
\newcommand{\sw}{\sin\theta_W}
\newcommand{\tw}{\tan\theta_W}
\def\cww{\cos^2\theta_W}
\def\sww{\sin^2\theta_W}
\def\tww{\tan^2\theta_W}

\def\noi{\noindent}
\def\nn{\noindent}

\def\sinb{\sin\beta}
\def\cosb{\cos\beta}
\def\sinbb{\sin (2\beta)}
\def\cosbb{\cos (2 \beta)}
\def\tgb{\tan \beta}
\def\tgbt{$\tan \beta\;\;$}
\def\tgbsq{\tan^2 \beta}
\def\sel{\tilde{e}_L}
\def\ser{\tilde{e}_R}
\def\msel{m_{\sel}}
\def\mser{m_{\ser}}


\def\mchi{m_\chi^+}
\def\neuto{\tilde{\chi}_1^0}
\def\mneuto{m_{\tilde{\chi}_1^0}}
\def\neutt{\tilde{\chi}_2^0}
\def\neutth{\tilde{\chi}_3^0}

\def\mstau{m_{\tilde\tau}}
\def\msne{m_{\tilde\nu}}
\def\mh{m_h}
\def\sinb{\sin\beta}
\def\cosb{\cos\beta}
\def\sinbb{\sin (2\beta)}
\def\cosbb{\cos (2 \beta)}
\def\tgb{\tan \beta}
\def\tgbt{$\tan \beta\;\;$}
\def\tgbsq{\tan^2 \beta}
\def\sinal{\sin\alpha}
\def\cosal{\cos\alpha}
\def\stop{\tilde{t}}
\def\sto{\tilde{t}_1}
\def\stt{\tilde{t}_2}
\def\stl{\tilde{t}_L}
\def\str{\tilde{t}_R}
\def\msto{m_{\sto}}
\def\mstosq{m_{\sto}^2}
\def\mstt{m_{\stt}}
\def\msttsq{m_{\stt}^2}
\def\mt{m_t}
\def\mtsq{m_t^2}
\def\sint{\sin\theta_{\stop}}
\def\sintt{\sin 2\theta_{\stop}}
\def\cost{\cos\theta_{\stop}}
\def\sintsq{\sin^2\theta_{\stop}}
\def\costsq{\cos^2\theta_{\stop}}
\def\mqtt{\M_{\tilde{Q}_3}^2}
\def\mutt{\M_{\tilde{U}_{3R}}^2}
\def\sbottom{\tilde{b}}
\def\sbo{\tilde{b}_1}
\def\sbt{\tilde{b}_2}
\def\sbl{\tilde{b}_L}
\def\sbr{\tilde{b}_R}
\def\msbo{m_{\sbo}}
\def\msbosq{m_{\sbo}^2}
\def\msbt{m_{\sbt}}
\def\msbtsq{m_{\sbt}^2}
\def\mt{m_t}
\def\mtsq{m_t^2}
\def\selectron{\tilde{e}}
\def\seo{\tilde{e}_1}
\def\set{\tilde{e}_2}
\def\sel{\tilde{e}_L}
\def\ser{\tilde{e}_R}
\def\mseo{m_{\seo}}
\def\mseosq{m_{\seo}^2}
\def\mset{m_{\set}}
\def\msetsq{m_{\set}^2}
\def\msel{m_{\sel}}
\def\mser{m_{\ser}}
\def\me{m_e}
\def\mesq{m_e^2}
\def\snu{\tilde{\nu}}
\def\snue{\tilde{\nu_e}}
\def\set{\tilde{e}_2}
\def\snul{\tilde{\nu}_L}
\def\msnue{m_{\snue}}
\def\msnuesq{m_{\snue}^2}
\def\smuon{\tilde{\mu}}
\def\smul{\tilde{\mu}_L}
\def\smur{\tilde{\mu}_R}
\def\msmul{m_{\smul}}
\def\msmulsq{m_{\smul}^2}
\def\msmur{m_{\smur}}
\def\msmursq{m_{\smur}^2}
\def\stau{\tilde{\tau}}
\def\stauo{\tilde{\tau}_1}
\def\staut{\tilde{\tau}_2}
\def\staul{\tilde{\tau}_L}
\def\staur{\tilde{\tau}_R}
\def\mstauo{m_{\stauo}}
\def\mstauosq{m_{\stauo}^2}
\def\mstaut{m_{\staut}}
\def\mstautsq{m_{\staut}^2}
\def\mtau{m_\tau}
\def\mtausq{m_\tau^2}
\def\gluino{\tilde{g}}
\def\mgluino{m_{\tilde{g}}}
\def\mchi{m_\chi^+}
\def\neuto{\tilde{\chi}_1^0}
\def\mneuto{m_{\tilde{\chi}_1^0}}
\def\neutt{\tilde{\chi}_2^0}
\def\mneutt{m_{\tilde{\chi}_2^0}}
\def\neutth{\tilde{\chi}_3^0}
\def\mneutth{m_{\tilde{\chi}_3^0}}
\def\neutf{\tilde{\chi}_4^0}
\def\mneutf{m_{\tilde{\chi}_4^0}}
\def\chargop{\tilde{\chi}_1^+}
\def\mchargo{m_{\tilde{\chi}_1^+}}
\def\chargtp{\tilde{\chi}_2^+}
\def\mchargt{m_{\tilde{\chi}_2^+}}
\def\chargom{\tilde{\chi}_1^-}
\def\chargtm{\tilde{\chi}_2^-}
\def\bino{\tilde{b}}
\def\wino{\tilde{w}}
\def\photino{\tilde{\gamma}}
\def\zino{tilde{z}}
\def\sdowno{\tilde{d}_1}
\def\sdownt{\tilde{d}_2}
\def\sdownl{\tilde{d}_L}
\def\sdownr{\tilde{d}_R}
\def\supo{\tilde{u}_1}
\def\supt{\tilde{u}_2}
\def\supl{\tilde{u}_L}
\def\supr{\tilde{u}_R}
\def\mh{m_h}
\def\mht{m_h^2}
\def\MH{M_H}
\def\MHt{M_H^2}
\def\MA{M_A}
\def\MAt{M_A^2}
\def\MHp{M_H^+}
\def\MHm{M_H^-}
\def\epem{e^+e^-}
\def\siginv{\sigma_{\gamma+inv}}
\def\gmuon{$(g-2)_\mu$}
\def\r12{r_{12}}
\def\bsgamma{b\ra s\gamma}

\pagestyle{empty}
\begin{center}
{\large {\bf {\em A lower limit on the neutralino mass in the MSSM with non-universal
gaugino masses.}}}
\vspace{1cm}

\begin{tabular}[t]{c}

{\bf G.~B\'elanger$^{1}$, F.~Boudjema$^{1}$, A. Pukhov$^{2}$, 
S. Rosier-Lees$^{3}$
}

\\
\\
1.{\large
LAPTH},
 {\it Chemin de Bellevue, B.P. 110, F-74941 Annecy-le-Vieux,
Cedex, France.}\\
2. Skobeltsyn Institue of Nuclear Physics, {\it Moscow State University, Moscow, Russia.}\\
3.{\large
LAPP},
 {\it Chemin de Bellevue, B.P. 110, F-74941 Annecy-le-Vieux,
Cedex, France.}\\

\end{tabular}
\end{center}

\centerline{ {\bf Abstract} } \baselineskip=14pt \noindent
{\small We discuss constraints on SUSY models with non-unified gaugino masses.
We concentrate on the slepton/gaugino sector and obtain
 a lower limit on the neutralino mass  combining direct limits,  indirect 
limits as well as  relic density measurements. 
}

\baselineskip=14pt


\section{Introduction}

In the framework of the mSUGRA model
which contains a rather limited number of parameters,  many detailed analysis on 
the constraints from various collider experiments, high precision  
lower energies
measurements as well as from  the point of view of dark matter have been 
performed \cite{baersugra}-\cite{fiorenza}.
However it is interesting to relax some of the
assumptions that go into  mSUGRA 
and strive for a more model independent
approach \cite{baernonuni}-\cite{olive}.

The main limits on the MSSM parameters come from LEP and Tevatron direct searches 
for SUSY particles and for the Higgs as well as from the muon anomalous 
moment \gmuon, the $b\ra s \gamma$ process and  the relic density of the lightest 
supersymmetric particle (LSP).
The LEP limits, the \gmuon~limit as well as, to a large extend,
the relic density, 
affect mainly the gaugino and slepton sector. 
On the other hand, the Tevatron is more sensitive to the 
coloured sector  while the 
Higgs mass lower limit   severely restricts the  
 low $\tan\beta$ region and is much  dependent on the squark sector
(especially $m_{\tilde t}$ and $A_t$). The branching ratio for $b\ra s \gamma$
also constrains the squark sector 
although it has some influence on the parameters of the gaugino/higgsino sector 
as it strongly favors $\mu>0$. 

A model independent approach 
with a manageable number of free parameters is then possible
when  one restricts oneself to the gaugino/sleptons sectors making only
mild assumptions about the reminder of the MSSM parameters.
Of particular interest is the lower limit on the neutralino mass
that can be obtained in a general model. Indeed a light neutralino
LSP opens the door for a sizeable branching fraction of the Higgs into invisible
thus reducing universally the branching fractions into the usual discovery
channels for a light Higgs, $h\ra \gamma\gamma$ or $h\ra b\bar{b}$ \cite{hinv}.
This possibility has triggered analyses by the LHC experiments to search
for the Higgs with a sizeable branching fraction into invisible
\cite{zeppenfeld}.

Here we concentrate solely on the lower limit on the neutralino mass 
in the general MSSM model. 
In particular we treat  the case where the  masses of $\tilde{l}_R$ and $\tilde{l}_L$ are not correlated. 
We also discuss the $\mu<0$ as well as the large $\tan\beta$ cases even though
 they do not lead to  the largest invisible Higgs width.

\section{MSSM parameters}

In the approach we are taking, the free parameters include the ones of the
 gaugino/higgsino sector as well as the parameters of the slepton sector. 
 We  consider two options, a model reminiscent of mSUGRA 
 models featuring  a common
 mass for the sleptons at the GUT scale (model M0) and a model 
 where the slepton masses are  not correlated (model LR).
 We impose universality among the generations. 
 Altogether we allow 6(7) free parameters in model M0 (LR):
 \beqn
 \tan\beta,M_1,M_2,\mu,A_l,m_0 \;\;\; (\msel,\mser)
  \eeqn
 Using the renormalization group equations 
 the masses at the weak scale can be related to the ones at the GUT scale, in model M0. For sleptons this can be done rather
independently of the other MSSM parameters \cite{hinv}.
\noi
We assume that all the squarks are heavy and that the pseudoscalar mass $M_A$ is large.
We will only consider
values of $\tan\beta> 5$ to make it easier  to satisfy the direct
limit $m_h>114 GeV$. 
To characterize the amount of non universality we  define the parameter
$r_{12}=\frac{M_1}{M_2}$.
For scans over parameter space, unlesss otherwise specified, we will consider
the range
\beqn
\label{scan}
5<\tan\beta<50,
 M_2<2~TeV,\;\;
  .001<\r12<.6,\;\;
  |\mu|<1~TeV,
   m_0(\msel,\mser)<1~TeV 
\eeqn
\noi
We will usually fix $A_l=0$ as most of the processes we will discuss are not
very sensitive to the exact value of this parameter for the sleptons.
For the squarks we take a value for $A_t$ that gives a large enough Higgs mass.

\section{Direct limits from LEP}

The direct limits from LEP on gauginos as well as on sleptons 
are relevant to obtain a lower bound  on the lightest neutralino
as the sleptons play an important role in the relic density
calculation. 
The LEP experiments obtain a  lower limit on the neutralino mass 
while  assuming  unified  gaugino masses at the GUT scale.
The constraint on the neutralino mass is basically derived
from the lower limit on the chargino mass obtained in the 
pair production process. The lower bound on the chargino 
rests near the kinematic limit,
$\mchi> 103.5$GeV
when sneutrinos are heavy, and drops to $\mchi>73$ GeV 
when   $75<\msne<85$GeV due to the destructive interference between the
 t- and s-channel contributions.
In a general MSSM, the charginos and neutralino  masses are 
uncorrelated and the lower limit on the neutralino mass weakens 
when $\r12<.5$. The production of  $\chi_1^0\chi_2^0(\chi_3^0)$
can be used to somewhat constrain the parameter space. 
In our scans we implemented
  the upper limit from the L3 experiment on these cross-sections \cite{L3_susy}.
For selectrons, a limit of 99.5GeV can be set on both $\sel$ as well as $\ser$
in the case of a light neutralino,  whereas  
 basically model independent
limits of $m_{\smuon} > 96$GeV  and  $m_{\stau} > 86$GeV  
can be reached \cite{lep_susy_WG}.

The radiative processes where
a photon is emitted in addition to a pair of invisibly decaying
supersymmetric particles  will contribute to the 
process $\epem\ra \gamma+invisible$ which has been searched for by
the LEP2 experiments. Such processes can be used to search for the lightest 
neutralino \cite{pandita_photino} or sneutrinos decaying to $\tilde\nu\ra\chi^0\nu$. 
The radiative processes
can also help  closing some loopholes in the LEP analyses
in the case of  charged sparticles 
that decay invisibly when they are nearly degenerate in mass with the LSP.
 Using calcHEP \cite{comphep}
we have computed all radiative processes involving sleptons and gauginos.

 For the lightest neutralino, after scanning over a wide range of parameters 
 in the MSSM, we found that the cross-section could reach $\sigma=50$fb for
  $m_0=100$GeV.
 This is below the value reached by LEP,
 approximately $\sigma< 0.2(.1)$pb for one(four) experiment(s). 
 However,
for sneutrinos, the cross-section for the radiative
process often exceeds these limits.
 For sleptons nearly degenerate with the LSP, we found 
 lower limits on the neutralino even more stringent than
 in the case without mass degeneracy, for example,
 $\mser\approx \mneuto> 56$GeV for  $\tan\beta=10,\mu>0$ in model M0.

\section{Indirect limits : relic density, \gmuon, $b\ra s\gamma$}

A MSSM model with a light neutralino
 must be consistent at least with the 
 upper limit on the amount of cold dark matter ($\Omega h^2< .3$).
Our calculations of the relic 
density is based on {\tt micrOMEGAs}, a program that calculates the relic 
density in the MSSM including all possible coannihilation channels \cite{cpc}. 
For the light neutralino masses under consideration,
 it is  the main annihilation channels that are most relevant, in particular
 annihilation  into a pair of light fermions. 
 Basically two diagrams contribute, s-channel Z (or Higgs) and
t-channel sfermion exchange. A light neutralino that is mainly a Bino couples
preferentially to right-handed sleptons, the ones that have the largest
hypercharge. To have a large enough annihilation rate (in order to bring down
the relic density below the upper limit allowed) one needs either a light
slepton or a mass close to $M_Z/2$. 
In the former case, the constraint from LEP plays an important role. 
In the heavy slepton case,  the coupling of the Z should 
be substantial, which requires that the neutralino
should have a certain Higgsino component. This means $\mu$ small, but still
consistent with the  chargino constraint.

Both the theoretical predictions and experimental results on the
muon anomalous magnetic moment have been refined on several occasions in the last year. At this conference, we presented results obtained using the bound
\beqn
\delta a_\mu=\left(23\pm 15_{|{\rm exp.}} \pm 14_{|{\rm
theo.}}\right) \;\times 10^{-10}.
\eeqn
Adding linearly the theoretical error
to  a $2\sigma$ experimental
error, this 
 translates into 
\beqn
-21 \;<\; \delta a_\mu\;\times 10^{10} \;<\; 67.
\label{g2}
\eeqn
Here, the  hadronic vacuum polarisation  is extracted after averaging the 
$e^+e^-$ and $\tau$ data \cite{g-2old}.
This value has been updated in the last few weeks 
with a more precise experimental result \cite{g-2exp} as well 
as new estimates of the hadronic vacuum polarisation \cite{g-2teubner}.
The allowed $2\sigma$ range (using $e^+e^-$ data alone) becomes
\beqn
-3 \;<\; \delta a_\mu\;\times 10^{10} \;<\; 67
\eeqn
To be very conservative and since many issues need to be clarified in particular
concerning the estimation of the  hadronic polarisation, we will still 
 discuss the limits presented in Eq.~\ref{g2}.  Note that in this case, the lower bound corresponds roughly to the  $5\sigma$ limit based on the most recent
results using   only $e^+e^-$ data \cite{g-2wells}.
In that scenario, the values  $\mu<0$, although severely constrained
at large $\tan\beta$ are not completely ruled out. 
Note that in general the sign of $\mu$ is strongly correlated to the one of $\delta a_\mu$, however cancellations between
 the chargino and the neutralino diagrams can change the relative sign 
of  $\delta a_\mu$ and $\mu$.  
When $\mu>0$,   only
scenarios with large $\tan\beta$, small $m_0$ and light neutralinos/charginos
are expected to exceed the upper limit.

The $b\ra s \gamma$
depends mostly on the squark and gaugino sector.
We  find that  for a given value of $\tan\beta$ it is always possible to find 
some values of $A_t$ that brings the branching ratio in the allowed 
range, 
\beqn
\label{bsgbound} 2.04 \times 10^{-4}< Br(b\ra s\gamma)\; < 4.42 \times 10^{-4}
\eeqn
over the full set of parameter space and that is also consistent with the Higgs
mass.

\section{Results} 

\noi
1) Heavy sleptons, $m_0=500$GeV.

In the case of heavy sleptons, the
main constraint on the neutralino arises from the relic density.
The contribution from sfermion exchange to neutralino annihilation should be
negligible.   Then in order to have a sufficient annihilation rate, which means sufficient
coupling to the Z, a certain amount of Higgsino component is necessary.
We scanned over the parameters $\r12,M_2,\mu,\tan\beta$ as specified in Eq.\ref{scan}. 
The minimum value for the mass is $\mneuto > 27 GeV$  
and occurs for $\r12<.2$. 
This lower bound is more or less independent of $\tan\beta$ and occurs for $|\mu|\approx 100$GeV. 
Improving the upper bound on the relic density would strengthen the lower limit
on $\mneuto$ by a few GeV's. However one cannot do much better
than $\approx 35GeV$. Indeed, when $\mneuto$ approaches $M_Z/2$, the effect of the Z
peak becomes so important that the relic density constraint is easily satisfied.
As $r_{12}$ increases, the direct limit from the chargino mass dominates.

\noi
2) Light sleptons, $\mu>0$
 
\begin{figure*}[htbp]
\begin{center}
\vspace{-1.7cm}
\includegraphics[width=15cm,height=9.5cm]{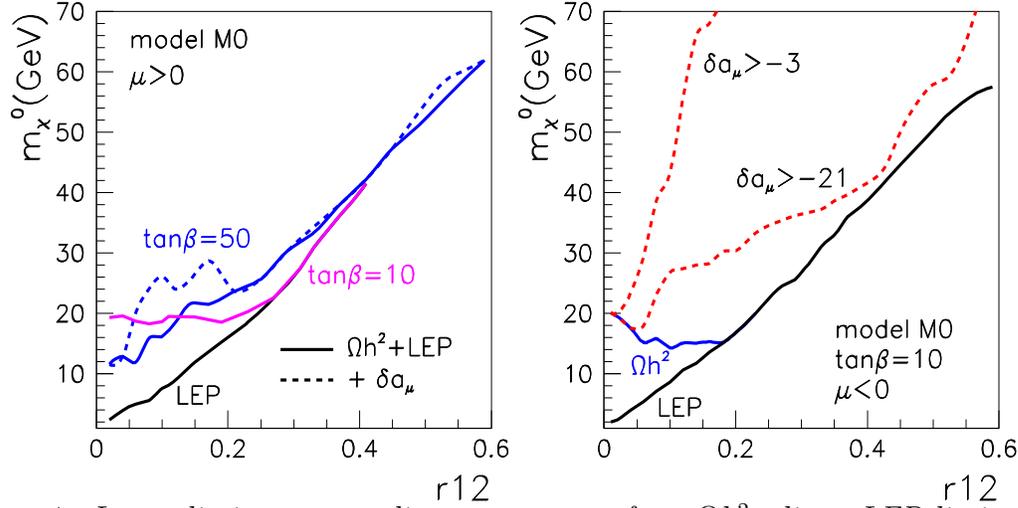}
\vspace{-1.9cm}
\caption{\label{tg10lower}{\em Lower limit on neutralino mass {\it vs}
$\r12$ 
from $\Omega h^2$,  direct LEP limits as well as $\delta a_\mu>-21(-3)\times10^{-10}$ for 
a) $\tan\beta=10,50$ and $\mu>0$ b) $\tan\beta=10,\mu<0$.}}
\end{center}
\end{figure*}

One expects the contribution from
t-channel sfermions to weaken the constraint from the relic density. 
Here the constraints on the mass of sfermions 
from LEP must be taken into account. 
 We show in Fig.\ref{tg10lower} the lower limit on the neutralino mass  
 as a function of $r_{12}$ after scanning over 
$M_2,\mu$ and $m_0$. 
The lower bound on the neutralino mass rests at 18GeV for $\tan\beta=10$
and $r_{12}< .2$. This lower bound increases with $r_{12}$ and follows
the limit from the chargino mass.
At larger values of $\tan\beta$, the relic density constraint is more severe 
except for $\r12< .1$. Furthermore one starts to marginally see  the impact of 
the \gmuon. 

\vspace{-.3cm}
\noi
3)Light sleptons, $\mu<0$.

\begin{figure*}[htbp]
\begin{center}
\vspace{-1.5cm}
\includegraphics[width=15cm,height=9.5cm]{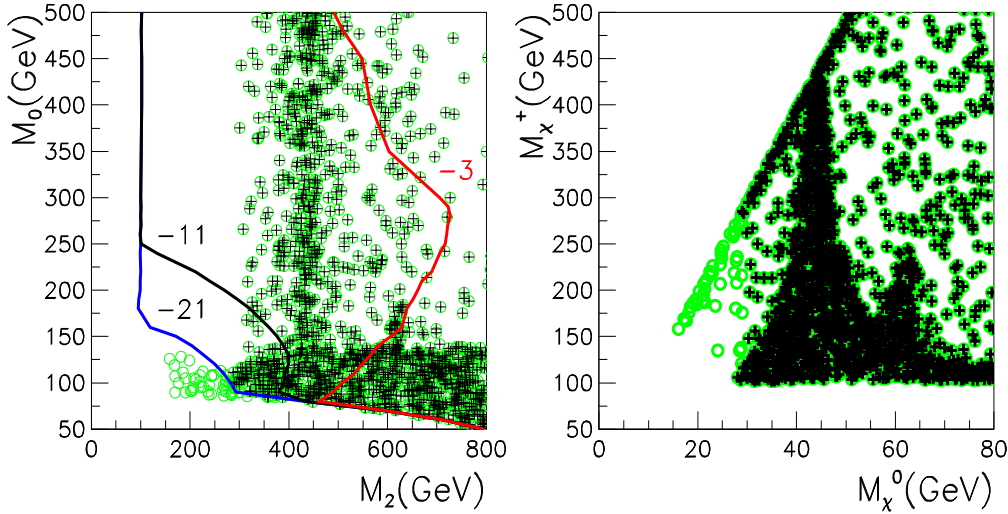}
\vspace{-1.5cm}
\caption{\label{tg10r1}{\em  Impact of the $\delta a_\mu=-21\times 10^{-10}$ 
constraint (crosses,dark grey) on the allowed region (circles,light grey)
 in the  a) $M_0-M_2$ plane 
b) $\mchi-\mneuto$ plane for $\mu<0,\r12=.1,\tan\beta=10$.
Contours for $\delta a_\mu=-21,-11,-3\times 10^{-10}$ are displayed.}}
\end{center}
\end{figure*}

The main difference with the case discussed above is that
here the \gmuon~ constraint plays an important role. 
To get some insight on the effect of combining various constraints we
first consider the special case, $\mu<0,\r12=.1$.
The free parameters are $M_2,m_0,\mu$.
As already mentioned it is in the  small $m_0$ region that one finds
the lightest neutralino. However, it is precisely in 
that region of the $m_0-M_1$ plane that one gets too large a contribution 
to $\delta a_\mu$ as displayed in Fig.\ref{tg10r1}  for $\tan\beta=10$.
Coupling the limit from $\delta a_\mu$ with the one from the relic density
then considerably strengthens the lower limit on the neutralino mass, 
$\mneuto>27$GeV. Furthermore one finds that 
the lightest neutralino allowed are necessarily accompanied by light charginos.
For larger values of $\tan\beta$,  
it becomes increasingly difficult to accommodate light neutralinos due to 
the  constraint from $\delta a_\mu$.

From a full scan over the parameter space for $\mu<0$ we find that
combining the three  constraints increases the lower bound on the neutralino
especially in the $\r12\approx .1-.4$ region (Fig.\ref{tg10lower}). 
When $M_1<<M_2$ cancellations between the neutralino and chargino contributions to $\delta a_\mu$ can occur thus making it possible to have $\mneuto\approx 20$GeV.  The impact of the \gmuon  is more drastic
for very large $\tan\beta$. These results hold for the  model M0. 

For model LR, allowing light $\tilde{e}_L$ has no significant impact 
on the relic density contribution which is  dominated by $\tilde{e}_R$. However 
 the \gmuon~ is much relaxed 
and basically one recovers more or less the results obtained without 
the   constraint from \gmuon.

\section{Conclusion}

Combining the upper limit from the relic density of neutralinos with direct limits
from LEP on gauginos and sleptons constrains the MSSM with non universal gaugino
masses. In particular a lower limit on the neutralino mass is obtained,
$\mneuto> 12-18$GeV for $M_1<<M_2$, leaving ample room for an invisibly decaying Higgs boson. 
 The implication of such models at linear colliders are dicussed in
 Ref.\cite{long}.

\noi {\bf Acknowledgments}

\noi
  This work was supported in part by the PICS-397.

\end{document}